\documentclass[namedreferences]{solarphysics}
\usepackage[hyperref,optionalrh,solaromanenum]{spr-sola-addons} % For Solar Physics 
\usepackage{graphicx}                    % For eps figures, newer & more powerfull
\usepackage{amssymb}                    % useful mathematical symbols
\usepackage{color}                       % For color text: \color command
\usepackage{breakurl}                         % For breaking URLs easily trough lines
\begin{document}

\begin{article}

\begin{opening}

\title{Comparison of Latitude Distribution and Evolution of Even and Odd Sunspot Cycles}

%%%%%%%%%%%%%%%%%%%%%%%%%%%%%%%%%%%%%%%%%%%%%%%%%%%
%% Authors Names
%
\author[addressref={aff1},corref,email={jouni.j.takalo@oulu.fi}]{\inits{J.J.}\fnm{Jouni}~\lnm{Takalo}}

\institute{$^{1}$ Space physics and astronomy research unit, University of Oulu, POB 3000, FIN-90014, Oulu, Finland\\
\email{jouni.j.takalo@oulu.fi}
}
%\address[id=aff1]{ReSoLVE Centre of Excellence, Space Climate research unit, University of Oulu,
%POB 3000, FIN-90014, Oulu, Finland}
%\date{Received: }
%%%%%%%%%%%%%%%%%%%%%%%%%%%%%%%%%%%%%%%%%%%%%%%%%%%
%% Runningheads
%
\runningauthor{J. Takalo}
\runningtitle{Latitude Distribution of Sunspot Cycles}

%%%%%%%%%%%%%%%%%%%%%%%%%%%%%%%%%%%%%%%%%%%%%%%%%%%
%% Affilations 
%% id shold be the same with \author addressref value.
%\address[id={}]{}

%%%%%%%%%%%%%%%%%%%%%%%%%%%%%%%%%%%%%%%%%%%%%%%%%%%
%%% Abstract 
\begin{abstract}
We study the latitudinal distribution and evolution of sunspot areas from Solar Cycles 12\,--\, Solar Cycles 23 (SC12-SC23) and sunspot-groups of from Solar Cycles 8\,--\,Solar Cycles 23 (SC8-SC23) for even and odd cycles. The Rician distribution is the best-fit function for both even and odd sunspots group latitudinal occurrence. The mean and variance for even northern/southern butterfly wing sunspots are 14.94/14.76 and 58.62/56.08, respectively, and the mean and variance for odd northern/southern wing sunspots are 15.52/15.58 and 61.77/58.00, respectively. Sunspot groups of even cycle wings are thus at somewhat lower latitudes on the average than sunspot groups of the odd cycle wings, i.e., about 0.6 degrees for northern hemisphere wings and 0.8 degrees for southern hemisphere wings.

The spatial analysis of sunspot areas between SC12-SC23 shows that the small sunspots are at lower solar latitudes of the sun than the large sunspots for both odd and even cycles, and also for both hemispheres.

Temporal evolution of sunspot areas shows a lack of large sunspots after four years (exactly between 4.2\,--\,4.5 years), i.e., about 40\% after the start of the cycle, especially for even cycles. This is related to the Gnevyshev gap and is occurring at the time when the evolution of the average sunspot latitudes cross about 15 degrees.
The gap is, however, clearer for even cycles than odd ones. Gnevyshev gap divides the cycle into two disparate parts: the ascending phase/cycle maximum and the declining phase of the sunspot cycle. 
\end{abstract}

%%%%%%%%%%%%%%%%%%%%%%%%%%%%%%%%%%%%%%%%%%%%%%%%%%%
%% Keywords
%

\end{opening}
%-------------------------------------------------

%%%%%%%%%%%%%%%%%%%%%%%%%%%%%%%%%%%%%%%%%%%%%%%%%%%
%% Sections

\section{Introduction}

It has been known almost for two hundred years that the occurrence of sunspots is cyclic, although not strictly periodic. The length of the sunspot cycle (SC) has varied between 9.0 and 13.7 years. The shape of the sunspot cycle has also changed. Waldmeier noticed the asymmetry of sunspot cycles, with the ascending phase being typically shorter than the declining phase, and that there is an anti-correlation between cycle amplitude and the length of the ascending phase of the cycle \citep{Waldmeier_1935, Waldmeier_1939}.

\cite{Gnevyshev_1967} suggested that the solar cycle is characterized by two periods of activity and these lead to a double peak with the so-called Gnevyshev-gap (GG) in between \citep{Gnevyshev_1977}. \cite{Feminella_1997} studied the long-term behaviour of several solar-activity parameters and found that maxima occur at least twice: first, near the end of the rising phase and then in the early years of declining phase. \cite{Norton_2010} analysed the sunspot cycle double peak, and the Gnevyshev gap between them, to determine if the double-peak is caused by averaging of the two hemispheres that are out of phase \citep{Temmer_2006}. They, however, confirmed previous findings that the Gnevyshev gap is a phenomenon that occurs separately in each hemisphere and is not due to a superposition of sunspot indices from hemispheres slightly out of phase.

Most of the even-odd cycle comparisons have concentrated on the mutual strength of preceding cycles. These are referred as the so-called
Gnevyshev-Ohl rule, which is an expression of the general 22-year variation of cycle amplitudes and intensities, according to which even cycles are, on average, about 10\,--\,15\% lower than the following odd cycles \citep{Mursula_2001}. There have, however, been some exceptions to this rule, the last one happened between cycle pair SC22-SC23 \citep{Javaraiah_2012, Javaraiah_2016}.

Another common subject has been the northern-southern asymmetry in solar sunspots and other activity; see some of the recent publications \citep{Temmer_2006, Carbonell_2007, Li_2009, Chang_2012, Javaraiah_2016, Badalyan_2017}. Less attention has been paid to statistical distributions of the sunspots, sunspot groups, and areas. There have, however, been some studies about the spatial (latitudinal) distribution of sunspots and sunspot groups using slightly different statistical methods \citep{Li_2003, Ivanov_2011, Chang_2012, Munoz-Jaramillo_2015, Leussu_2016, Leussu_2017, Mandal_2017}.

\cite{Li_2003} analysed the latitudinal distribution of sunspot groups over a solar cycle. They found that although individual sunspot groups of a solar cycle emerge randomly at any middle and low latitude, the whole latitudinal distribution of sunspot groups of the cycle is not stochastic and, in fact, can be represented by a probability density function of the Gamma- distribution type having a maximum probability at about 15.5 degrees of heliographic latitude.
\cite{Ivanov_2011} use in their analysis the relative latitude of sunspot groups, i.e. $\phi_{rel}=\phi-\phi_{o}$, where $\phi_{o}$ is the mean heliographic latitude and $\phi$ is the actual latitude. They show that the latitude distributions of sunspots for a given year can be approximately described by the normal distribution, with its variance being a linear function of the current level of solar activity. Thus, the latitude size of the spotted zone increases with increasing activity.
\cite{Chang_2012} studied the latitudinal distribution of sunspots between 1874\,--\,2009 by calculating the area-weighted sunspot mean value for each month (he calls it center-of-latitude, COL). He showed that the COL for both northern and southern hemisphere can be fitted with a double Gaussian distribution function.

\cite{Munoz-Jaramillo_2015} use 11 different sunspot group, sunspot, and active region databases to characterize the area and flux distributions of photospheric magnetic structures. They plot the fraction of the areas/fluxes per unit area as a function of their area/flux (in millionths of hemisphere, MH), and fit different distributions, power-law, log-normal, exponential and Weibull distributions to the these databases. They found that Weibull (six databases) and log-normal (four databases) are best-fit distributions, but none can explain the small areas (weak fluxes) and large areas (strong fluxes). They conclude that a composite distribution with Weibull for small areas (weak fluxes) and log-normal for large areas (strong fluxes) with transition regions in between can explain best each of their databases.
\cite{Leussu_2016} studied especially the latitude evolution and the timing of the sunspot groups in butterfly wings by characterizing three different categories: the latitude at which the first sunspot groups appear, the maximum latitude of sunspot group occurrence in each wing, and the latitude at which the last sunspot group appears. They derived several statistical measures based on these variables.
\cite{Mandal_2017} found that the latitude distribution of sunspots do not statistically follow a Gaussian distribution. Furthermore, they found that the distribution parameters, the central latitude, width and height show some hemispheric asymmetries and significant positive correlations with the cycle strengths.

In this study we analyse the spatial distribution and evolution of sunspot areas of Solar Cycles (SC) 12\,--\,SC23 and sunspot groups of SC8\,--\,SC23 for even and odd cycles, and find the best-fit probability density functions (PDF) for these distributions. This article is organized as follows. Section 2 presents the data and methods used. In Section 3 we study distributions and PDF fits of sunspot groups for even and odd solar cycles. In Sections 4 we discuss the latitudinal distribution of sunspot area sizes and migration of the sunspots over the solar cycle. We give our conclusions in Section 5.

\section{Data and Methods}

\subsection{Sunspot Group Data}

\begin{figure}
	\centering
	\includegraphics[width=1\textwidth]{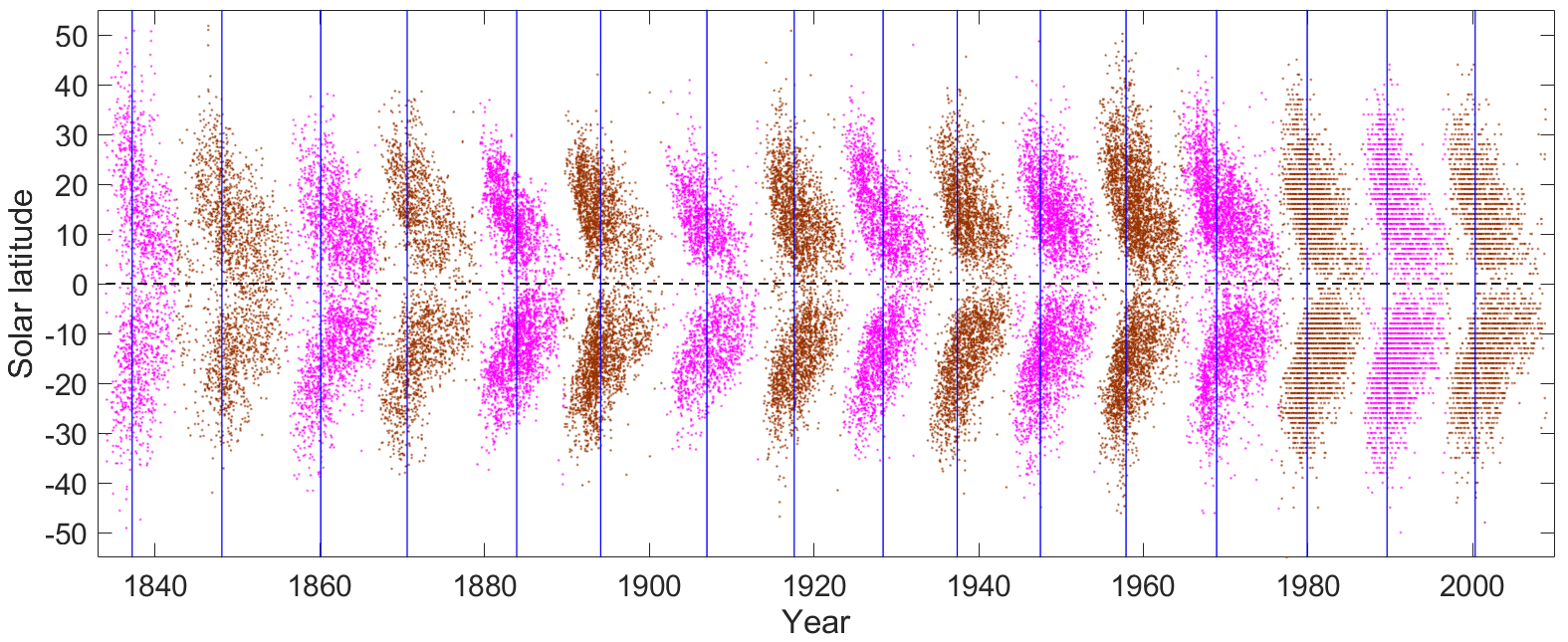}
		\caption{Butterfly pattern of the sunspot groups used in the analysis. The vertical lines correspond to cycle maxima.}
		\label{fig:Leussu_data}
\end{figure}

In the distribution analysis we use the data set of sunspot groups in southern and northern butterfly wings for cycles SC8-SC23 by Leussu \textit{et al}. \citep{Leussu_2017}. This data includes the time and latitude for sunspot groups for cycles SC8-SC23 and is seen as a butterfly pattern in Figure \ref{fig:Leussu_data}. Note, that the cycles SC21-SC23 have only integer degree resolution in latitudes.

\subsection{Sunspot Area Sizes}

In the aerial sunspot analysis we use the database of the Royal Observatory, Greenwich - USAF/NOAA Sunspot Data for the years 1874\,--\,2016 (https://\newline solarscience.msfc.nasa.gov/greenwch.shtml). This database contains, among others, time, latitude, and area size (in millionths of solar hemisphere, MH) for individual sunspots for cycles SC12-SC23.

\subsection{Distributions}

\subsubsection{Rayleigh Distribution}

The traditional definition for a normal distribution probability density function (PDF) is
\begin{equation}
f\,(x\,;\,\mu,\sigma^{2}) = \frac{1}{\sqrt{2\pi\,\sigma^{2}}} e^{ - \frac{\left({x-\mu }\right)^{2}}{2\,\sigma^{2}}} ,
\end{equation}
where $\mu$ is the mean (or expectation) value and $\sigma^{2}$ is the variance. Let us consider a bivariate distribution (U,V), where U and V are normal, independent distributions with zero mean. Then, we have
\begin{equation}
f_{U}\,(u\,;\,\sigma^{2}) = \frac{1}{\sqrt{2\pi\,\sigma^{2}}} e^{ - \frac{u^{2}}{2\,\sigma^{2}}} ,
f_{V}\,(v\,;\,\sigma^{2}) = \frac{1}{\sqrt{2\pi\,\sigma^{2}}} e^{ - \frac{v^{2}}{2\,\sigma^{2}}}.
\end{equation}
Let us define $X = \sqrt{U^{2}+V^{2}}$. Because $f_{U}$ and $f_{V}$ are perpendicular, their joint bivariate normal distribution, X, has a cumulative distribution
\begin{equation}
	F_{X}\,(x;\sigma^{2}) = \int\int_{D_{x}}\,f_{U}\,(u;\sigma^{2})\,f_{V}\,(v;\sigma^{2})\,dA ,
\end{equation}
where $D_{x}$ is the disc $D_{x}=\left\{\sqrt{u^{2}+v^{2}}<x\right\}$.
Integration is easier in polar coordinates, where $dA=r\,dr\,d\theta$. Writing the double integral in polar coordinates we get
\begin{equation}
	F_{X}\,(x;\sigma^{2}) = \frac{1}{2\pi\,\sigma^{2}}\int^{2\pi}_{0}\int^{x}_{0}\,r e^{ - \frac{r^{2}}{2\,\sigma^{2}}}\,dr\,d\theta = \frac{1}{\sigma^{2}}\int^{x}_{0}\,r e^{ - \frac{r^{2}}{2\,\sigma^{2}}}\,dr .
\end{equation}
While the probability function is the derivative of the cumulative function, we finally get
\begin{equation}
	f_{X}\,(x;\sigma^{2}) = \frac{d}{dx}\,F_{X}\,(x;\sigma^{2}) = \frac{x}{\sigma^{2}}e^{-\frac{x^{2}}{2\,\sigma^{2}}} , x\geq0 ,
\end{equation}
which is the Rayleigh PDF.

%\subsubsection{Weibull distribution}

%The probability density function of Weibull distribution is 
%\begin{equation}
%	f_{X}\,(x;\lambda,\beta) = \frac{\beta}{\lambda}\,\left(\frac{x}{\lambda}\right)^{\beta-1} %e^{-\left(\frac{x}{\lambda}\right)^{\beta}}\!, x\geq0
%\end{equation}

%It is evident that if $\beta$=2 we get

%\begin{equation}
%	f_{X}\,(x;\lambda,2) = \frac{2x}{\lambda^{2}} e^{-\left(\frac{x}{\lambda}\right)^{2}}\!, x\geq0  ,
%\end{equation}
%which is probability function of Rayleigh distribution, if $\lambda^{2}=2\sigma^{2}$ \citep{Forbes_2011}.

\subsubsection{Rician Distribution}

The Rice distribution, Rician distribution, is the probability distribution related to the Rayleigh distribution but with non-zero mean \citep{Rice_1945}. The PDF of the Rician distribution is defined as

\begin{equation}
f_{Z}\left(z\right)=\frac{z}{\sigma^{2}}\exp\left(\frac{-\left(z^{2}+\nu^{2}\right)}{\sigma^{2}}\right)I_{0}\left(\frac{z\nu}{\sigma^{2}}\right) , r\geq0 ,
\end{equation}
where $I_{0}$ is the modified Bessel function of the first kind with zeroth order \citep{Rice_1945,Taluktar_1991}. Note, that for $\nu=0$ we get the Rayleigh distribution. Note also, that $\nu$ and $\sigma^{2}$ are not the mean and variance of the Rician distribution, but are related to the bivariate Gaussian distribution 
$x \sim \mathcal{N}(\nu\,cos\left(\theta\right),\,\sigma^{2})$ and $y \sim \mathcal{N}(\nu\,sin\left(\theta\right),\,\sigma^{2})$ such that $z=\sqrt{x^{2}+y^{2}}$. The mean and variance of the Rician distribution are defined by

\begin{equation}
\mu = \sigma\sqrt{\pi/2}L_{1/2}\left(-\frac{\nu^{2}}{2\sigma^{2}}\right)
\end{equation}
and
\begin{equation}
var = 2\sigma^{2}+\nu^{2}-\frac{\pi\sigma^{2}}{2}\,L^{2}_{1/2}\left(-\frac{\nu^{2}}{2\sigma^{2}}\right) ,
\end{equation}
respectively. Here $L_{1/2}\left(\cdot\right)$ denotes a Laguerre polynomial, which can be presented as

\begin{equation}
L_{1/2}\left(\xi\right) = \exp^{1/2}\left[\left(1-\xi\right)I_{0}\left(-\frac{\xi}{2}\right)-xI_{1}\left(-\frac{\xi}{2}\right)\right] .
\end{equation}
% \section{}%\label{s:?} 
\subsection{Statistical Methods}

\subsubsection{Negative Log-Likelihood}

The likelihood function $L\left(\theta\right)$ is defined as
\begin{equation}
	L\left(\theta\right) = \prod^{n}_{i=1}f_{\theta}\left(x_{i}\right) ,
\end{equation}
if variables $x_{i}$ are independent and from the same distribution $f_{\theta}$. The set of parameters $\theta$ of the distribution, which maximizes $L\left(\theta\right)$ is called a maximum likelihood estimator (MLE) and is denoted as $\theta_{L}$. It is often easier to maximize the log-likelihood function, $log\,L\left(\theta\right)$, and since the (natural) logarithmic function is monotonically increasing, the same value maximizes both $L\left(\theta\right)$ and $log\,L\left(\theta\right)$. Because the log-likelihoods here are always negative, we calculate instead the minimum value for negative log-likelihood (NLogL) \citep{Forbes_2011}.

\subsubsection{Two-Sample T-Test}

The two-sample T-test for equal mean values is defined as follows. The null hypothesis assumes that the means of the samples are equal, i.e. $\mu_{1}=\mu_{2}$. Alternative hypothesis is that $\mu_{1}\neq\mu_{2}$. The test statistic is calculated as
\begin{equation}
T = \frac{\mu_{1} - \mu_{2}}{\sqrt{{s^{2}_{1}}/N_{1} + {s^{2}_{2}}/N_{2}}} ,
\end{equation}
where $N_{1}$ and $N_{2}$ are the sample sizes, $\mu_{1}$ and $\mu_{1}$ are the sample means, and $s^{2}_{1}$ and $s^{2}_{2}$ are the sample variances. If the sample variances are assumed equal, the formula reduces to
\begin{equation}
T = \frac{\mu_{1} - \mu_{2}} {s_{p}\sqrt{1/N_{1} + 1/N_{2}}} ,
\end{equation}
where
\begin{equation}
s_{p}^{2} = \frac{(N_{1}-1){s^{2}_{1}} + (N_{2}-1){s^{2}_{2}}} {N_{1} + N_{2} - 2} .
\end{equation}
The rejection limit for two-sided T-test is $\left|T\right| > t_{1-\alpha/2,\nu}$, where $\alpha$ denotes significance level and $\nu$ degrees of freedom. The values of $t_{1-\alpha/2,\nu}$ are published in T-distribution tables \citep{Snedecor_1989, Krishnamoorthy_2006, Derrick_2016}.  Now, if the value of p$<\alpha=0.05$, the significance is at least 95\%, and if p$<\alpha=0.01$, the significance is at least 99\%.

\section{Latitudinal Distribution of Sunspot Groups for Even and Odd Cycles}

\begin{figure}
	\centering
	\includegraphics[width=1.0\textwidth]{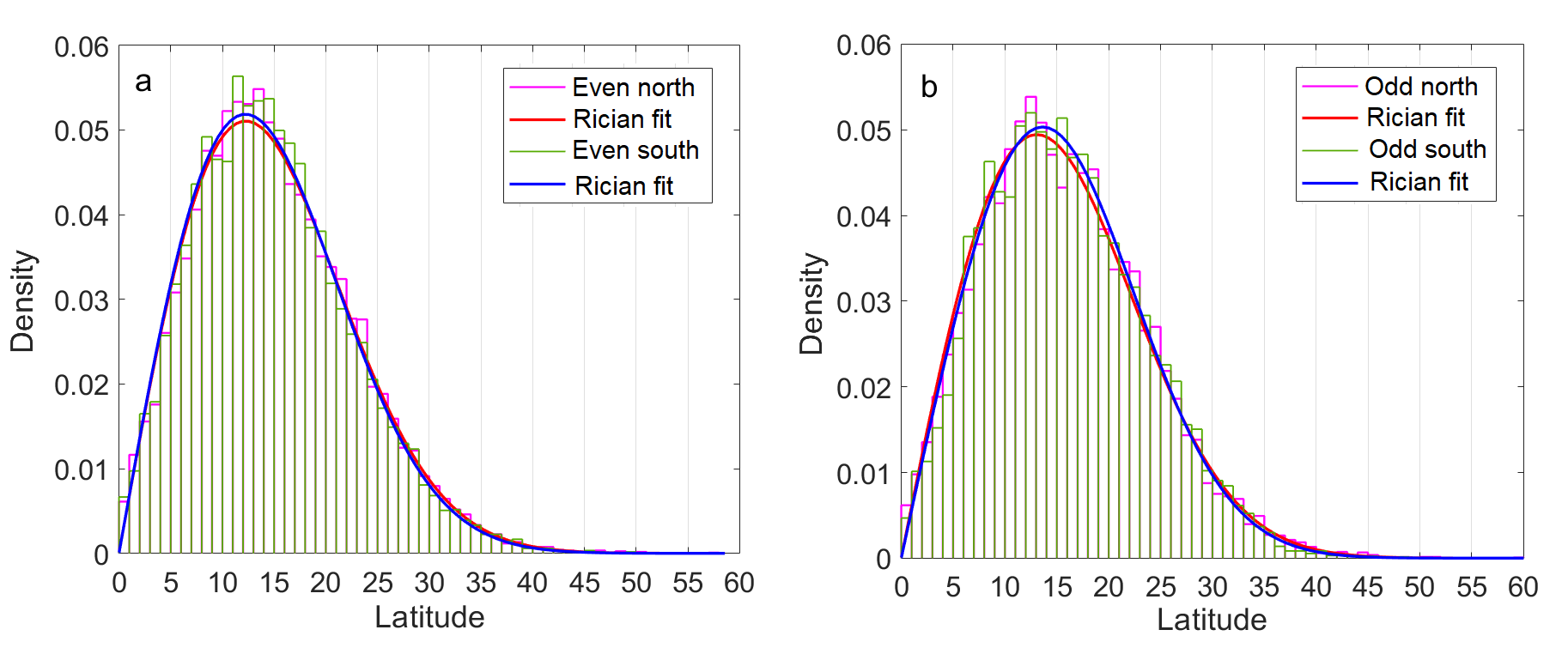}
		\caption{a) The latitudinal probability density of northern and southern (absolute value) hemisphere sunspot groups with their Rician distribution fits for even cycles in the range SC8\,--\,SC23. b) Same as a) but for odd cycles.}
		\label{fig:Latitude_distribution_even_odd_SPs}
\end{figure}

Figures \ref{fig:Latitude_distribution_even_odd_SPs}a and \ref{fig:Latitude_distribution_even_odd_SPs}b show the latitudinal distribution histograms of northern and southern (absolute value) hemisphere sunspot groups with their Rician distribution fits for even and odd cycles SC8\,--\,SC23, respectively. It is clear that the distributions are not normal because of the cutoff at zero and positive skewness. The NLogL of the Rician distribution fits for even northern/southern wing groups are 40761/40770, while NLogL for, e.g., lognormal distribution fits are 42347/42417. Almost as good fits as Rician are Rayleigh (NLogL = 40766/40788) and Weibull (40764/40774) distribution fits. The NLogL of Rician distribution fits for odd northern/southern wing sunspots are 46803/44526, while NLogL for lognormal distribution fits are 48770/46374. The Rician distribution parameters for even northern/southern wing sunspot are $\nu$= 8.262 (standard error, SE=0.581)/8.730 (SE=0.443) and $\sigma$= 10.338 (SE=0.234)/9.944 (SE=0.196) and the Rician distribution parameters for odd northern/southern wing sunspots are $\nu$= 9.822 (SE=0.334)/11.072 (SE=0.208) and $\sigma$= 10.162 (SE=0.167)/9.43713 (SE=0.119). The mean and variance for even northern/southern sunspot groups are 14.94/14.76 and 58.62/56.08, respectively, and the mean and variance for odd northern/southern sunspot groups are 15.52/15.58 and 61.77/58.00, respectively. This shows that the sunspot groups of even cycle wings are on the average somewhat lower latitudes than sunspot groups the odd cycle wings, i.e., about 0.6 degrees for northern hemisphere groups and 0.8 degrees for southern hemisphere groups. Two-sample T-test gives significance for the difference of the means between even and odd cycles with p$<10^{-9}$ and p$<10^{-16}$ for north sunspot groups and south sunspot groups, respectively. This means that significance is much better than 99\% in both cases.

\section{Distribution of Sunspot Areas}

\subsection{Latitudinal Distribution of Sunspot Areas}

\begin{figure}
	\centering
	\includegraphics[width=1.0\textwidth]{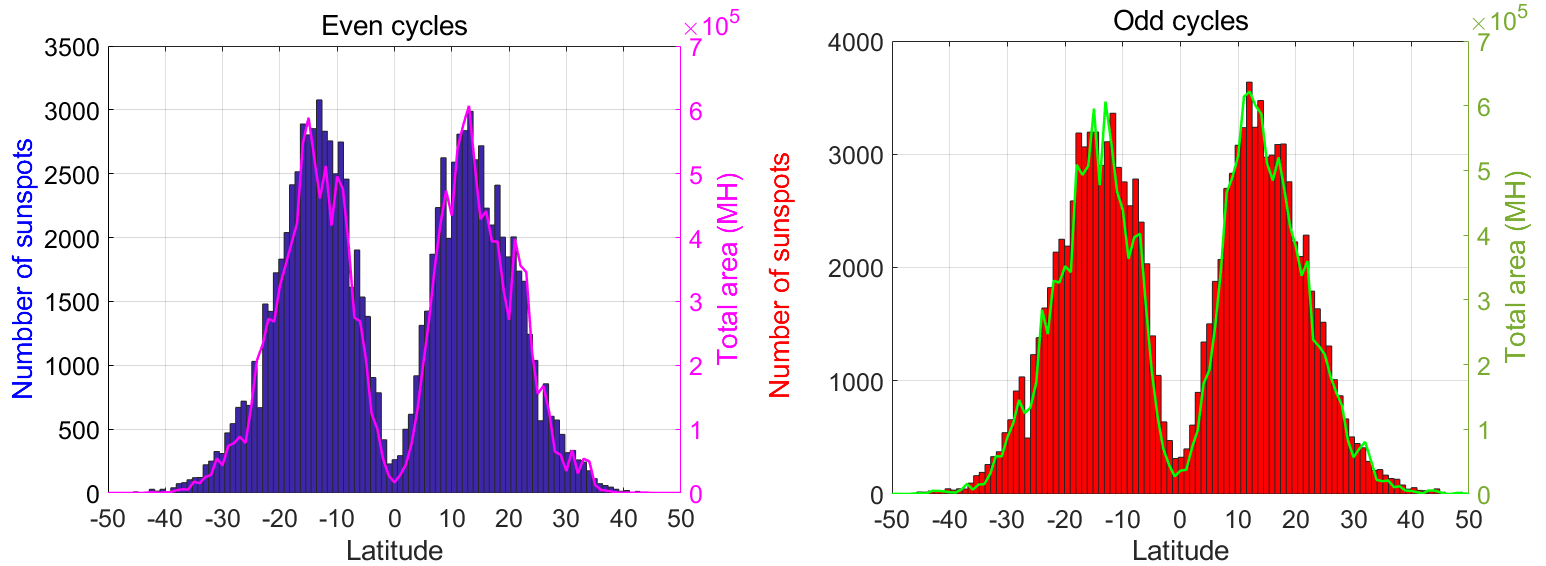}
		\caption{a) The number of all sunspots of even cycles  SC12\,--\,SC22 as histograms and their total areas (left vertical axis) for both northern and southern solar hemisphere. b) Same as a) but for all sunspots of odd cycles  SC13\,--\,SC23.}
		\label{fig:Even_odd_all_spatial_areas}
\end{figure}

Figures \ref{fig:Even_odd_all_spatial_areas}a and \ref{fig:Even_odd_all_spatial_areas}b show the density of all sunspots (independent of the area) as histograms and the total areas of the sunspots at each (integer) latitude  for even and odd cycles between SC12\,--\,SC23, respectively. The distribution and areal strength of the sunspots are quite similar but do not coincide in the low latitudes. It seems that there are smaller sunspots at low latitudes of the Sun for both odd and even cycles, and also for both hemispheres. Note, that the histograms are somewhat narrower for even cycles than for odd cycles in both hemispheres. The averages latitudes are 15.02 and -14.79 for the northern and southern hemisphere for even cycles, respectively, and 15.51 and -15.55 for the northern and southern hemisphere for odd cycles, respectively. Again the average latitudes are about 0.5 and 0.8 degrees lower for even cycles than odd cycles. The average sizes of sunspots are 165.5 and 159.1 MH for even and odd cycles, respectively.

\begin{figure}
	\centering
	\includegraphics[width=1.0\textwidth]{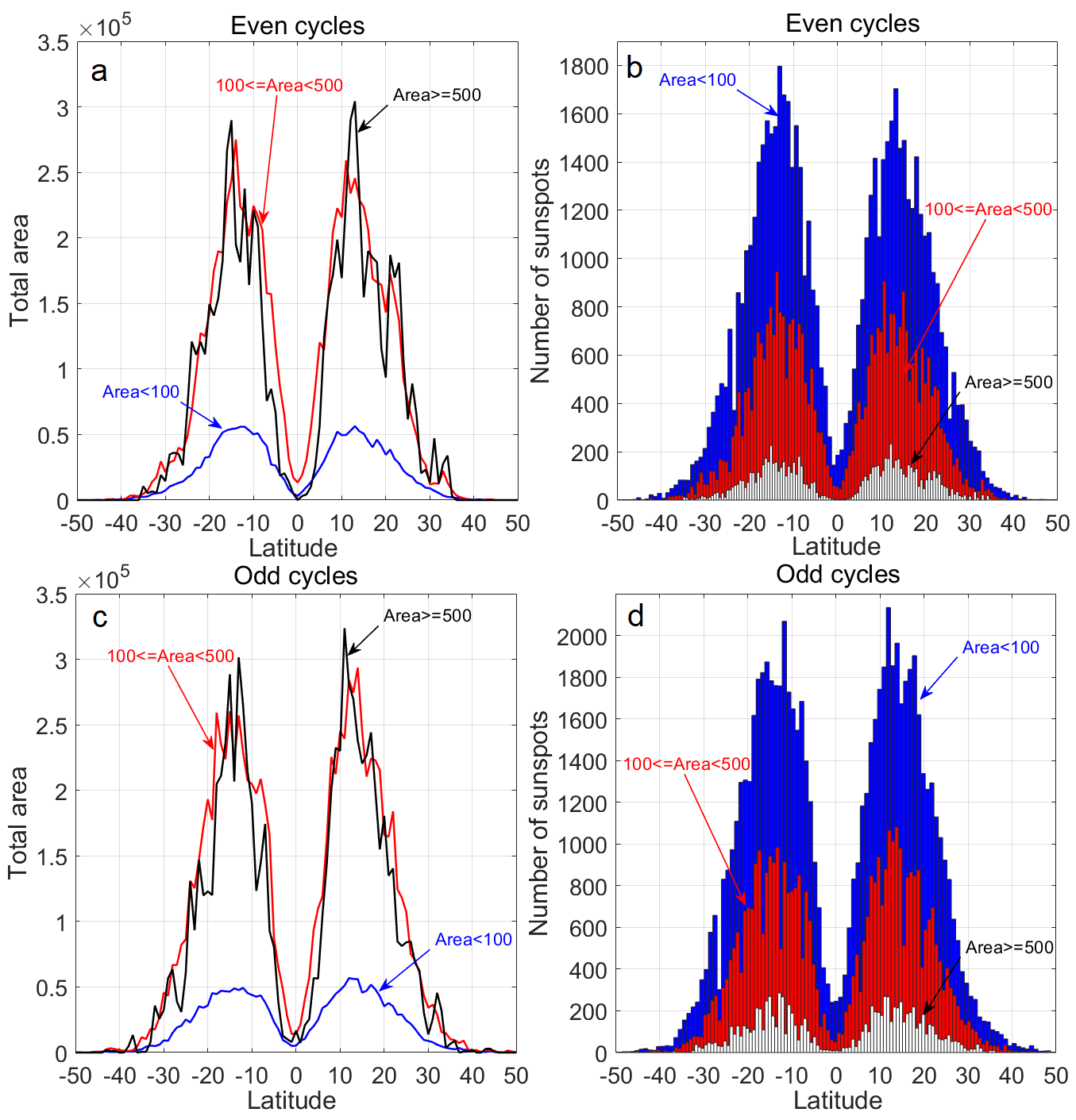}
		\caption{a) Total areas of large sunspots (area$\geq$ 500 MH), medium sunspots (100 $\leq$area$<$ 500 MH) and small sunspots (area$<$ 100 MH) of even cycles in the range SC12\,--\,SC22 and b)  the number of each size as histograms for both northern and southern solar hemisphere for even cycles. c) Same as a) but for sunspots of odd cycles in the range SC13\,--\,SC23 and d) same as c) but for odd cycles.}
		\label{fig:Even_odd_g1000_spatial_areas}
\end{figure}

Figures \ref{fig:Even_odd_g1000_spatial_areas}a and \ref{fig:Even_odd_g1000_spatial_areas}c show the total areas of large sunspots (area$\geq$ 500 MH), medium size sunspots (100 $\leq$area$<$ 500 MH) and small sunspots (area$<$ 100 MH) and number as histograms at each latitude in Figures \ref{fig:Even_odd_g1000_spatial_areas}b and \ref{fig:Even_odd_g1000_spatial_areas}d for even and odd cycles, respectively. It seems that the largest sunspots are located at latitudes between absolute values 10\,--\,25 degrees for both hemispheres and both even and odd cycles. There also seems to be some (quasi) periodicity in the locations of the large sunspots, especially for odd cycles. Note that for both hemispheres for even cycles and southern hemisphere for odd cycles there is a gap in the histogram of large sunspots around latitude $\pm$15 degrees.
The small sunspots are most abundant and spread from zero to 40 degrees but their total area is much smaller than the total areas of medium sized and large sunspots. The total areas of all sunspots are 1.78$\times10^{7}$ and 2.01$\times10^{7}$ MH for even and odd cycles of SC12\,--\,SC23, respectively.

\subsection{Temporal Distribution and Migration of Sunspots}

\begin{figure}
	\centering
	\includegraphics[width=1.0\textwidth]{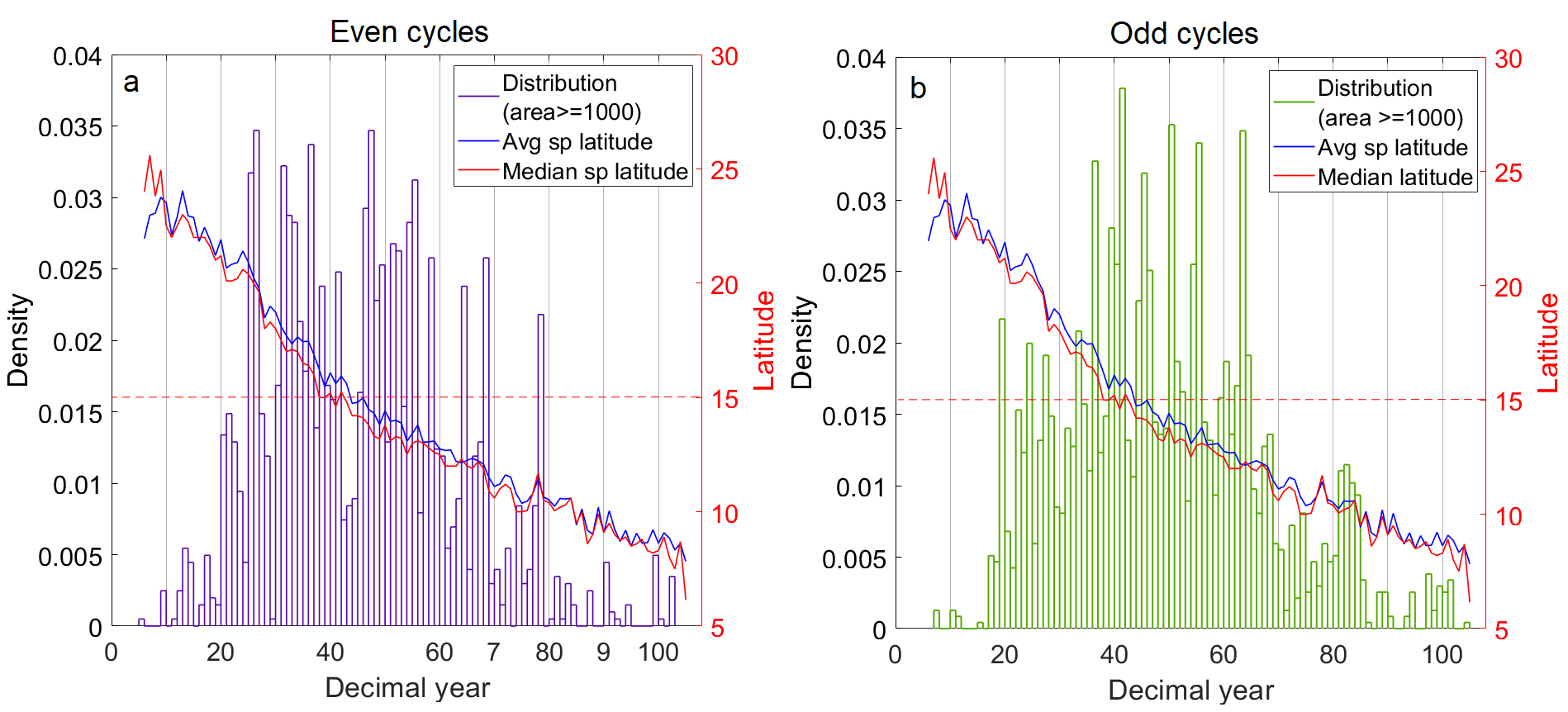}
	\caption{a) The density of the large sunspot distribution for even cycles in the range SC12\,--\,SC22 as a function of decimal year (unit=0.1 year). The blue and red curves show the mean and median latitude of these sunspots as a function of time (the latitude axis on the right). b) Same as a) but for large sunspots of odd cycles in the range SC13\,--\,SC23.}
		\label{fig:Even_odd_g1000_temporal_areas}
\end{figure}

Figures \ref{fig:Even_odd_g1000_temporal_areas}a and \ref{fig:Even_odd_g1000_temporal_areas}b show the density of very large sunspot distributions (areas equal or over 1000 MH) for even and odd cycles between SC12\,--\,SC23, respectively. The time-resolution for the figures is 0.1 year. In the same figures we show the average and median latitudes of all sunspots (right-axis) as a function of time for the whole period SC12\,--\,SC23 (absolute value is taken for southern hemisphere sunspots). The most conspicuous feature for the even cycles is the lack of large sunspots for four years (exactly between 4.2\,--\,4.5 years) after the start of the cycle. Two-sided T-test shows that this gap has significantly different mean value (at level 99\%) with p=0.0022 than the one-year period before and after the gap. Furthermore, the gap seems to locate at about the time when the evolution of the average sunspot latitude crosses about 15 degrees solar latitude. We believe that this is related to Gnevyshev-gap \citep{Gnevyshev_1967, Gnevyshev_1977, Storini_2003, Ahluwalia_2004, Bazilevskaya_2006, Norton_2010, Du_2015, Takalo_2018}. The distribution of odd cycles has only a small decrease in the distribution at 4.2\,--\,4.5 years, but it is insignificant according to two-sided T-test analysis (p=0.40).

\section{Conclusions}

We have analysed the spatial and temporal distribution of sunspot groups for even and odd cycles in the range SC8\,--\,SC23. The Rician distribution is the best-fit function for both even and odd sunspot-group latitudinal occurrence. The mean and variance for northern/southern wing sunspots are 14.94/14.76 and 58.62/56.08 for even cycles, respectively, and 15.52/15.58 and 61.77/58.00 for odd cycles, respectively. This shows that the sunspot groups of even cycle wings are on the average at somewhat lower latitudes than sunspot groups of the odd cycle wings, i.e. about 0.6 degrees for northern hemisphere wings and 0.8 degrees for southern hemisphere wings. Furthermore, the distributions for even cycles are slightly narrower than for odd cycles. The skewnesses for all (absolute value) even/odd cycle distributions are 0.589/0.500 and the kurtoses 3.303/3.147, which verify the above-mentioned results.

The spatial analysis of sunspot areas between SC12\,--\,SC23 shows that the small sunspots are at lower latitudes of the Sun than the large sunspots for both odd and even cycles, and also for both hemispheres. The total area of the sunspots is more symmetric for the hemispheres of the odd cycles than for those of the even cycles. The large area sunspots ($\geq$ 1000 MH) occur mainly between 10 and 25 degrees at both northern and southern hemisphere and for both even and odd cycles. For even cycles there is a gap at 15 (-15) degrees northern (southern) hemisphere latitudes in the occurrence of large sunspots, but for odd cycles the amount of large sunspots abruptly decreases after 15 (-15) degrees northern (southern) latitude.

The temporal evolution of sunspot areas shows a lack of large sunspots for four years (exactly between 4.2\,--\,4.5 years), i.e., about 40\% after the start of the cycle. This is related to the Gnevyshev gap and is consistent with the earlier result by \cite{Takalo_2018}. The significance level of this gap for even cycles is at least 95\% for all sunspots and still higher (~99\%) for large sunspots. Furthermore, the average of the sunspot areas in this interval for even cycles is 152 MH, while in the surrounding interval (year before and year after) it is 188 MH. We have also shown that this gap is temporally located at the same time as the average latitude of the migration of the sunspots crosses 15 (-15) degrees of northern (southern) heliographic latitude. For odd cycles the gap is narrower (4.2\,--\,4.3 years) and it is insignificant according to two-sample T-test for the means.

\begin{acknowledgements}

We acknowledge the financial support by the Academy of Finland to the ReSoLVE Centre of Excellence (project no. 272157). The author thanks prof. K. Mursula for useful discussions. The sunspot group data are downloaded from VizieR database (https://vizier.u-strasbg.fr/viz-bin/VizieR?-source=J/A+A/599/A131\&-to=3), and the sunspot area data from 
\newline RGO-USAF/
NOAA (https://solarscience.msfc.nasa.gov/greenwch.shtml). The dates of cycle maxima were obtained from from the National Geophysical Data Center (NGDC), Boulder, Colorado, USA (ftp.ngdc.noaa.gov). 

\end{acknowledgements}

\textbf{Disclosure of Potential Conflicts of Interest}

The author declares that there are no conflicts of interest.

%%% %%%%%%%%%%%%%%%%%%%%%%%%%%%%%%%%%%%%%%%%%%%%%%%%%%%%%%%%%%%
%% Bibliography
%
% Using BibTeX
%
\bibliographystyle{spr-mp-sola}
\bibliography{references_JT_SolPhys}  

\begin{thebibliography}{31}
% BibTex style file: spr-mp-sola.bst (nameyear), 2019-10-09
\ifx\bisbn     \undefined \def\bisbn  #1{ISBN #1}\fi
\ifx\binits    \undefined \def\binits#1{#1}\fi
\ifx\bauthor   \undefined \def\bauthor#1{#1}\fi
\ifx\batitle   \undefined \def\batitle#1{#1}\fi
\ifx\bjtitle   \undefined \def\bjtitle#1{\textit{#1}}\fi
\ifx\bvolume   \undefined \def\bvolume#1{\textbf{#1}}\fi
\ifx\byear     \undefined \def\byear#1{#1}\fi
\ifx\bissue    \undefined \def\bissue#1{#1}\fi
\ifx\bfpage    \undefined \def\bfpage#1{#1}\fi
\ifx\blpage    \undefined \def\blpage #1{#1}\fi
\ifx\burl      \undefined \def\burl#1{\textsf{#1}}\fi
\ifx\href      \undefined \def\href#1#2{\textsf{#2}}\fi
\ifx\betal     \undefined \def\betal{\textit{et al.}}\fi
\ifx\bctitle   \undefined \def\bctitle#1{#1}\fi
\ifx\beditor   \undefined \def\beditor#1{#1}\fi
\ifx\bbtitle   \undefined \def\bbtitle#1{\textit{#1}}\fi
\ifx\bedition  \undefined \def\bedition#1{#1}\fi
\ifx\bseriesno \undefined \def\bseriesno#1{\textbf{#1}}\fi
\ifx\blocation \undefined \def\blocation#1{#1}\fi
\ifx\bsertitle \undefined \def\bsertitle#1{\textit{#1}}\fi
\ifx\bsnm      \undefined \def\bsnm#1{#1}\fi
\ifx\bsuffix   \undefined \def\bsuffix#1{#1}\fi
\ifx\bparticle \undefined \def\bparticle#1{#1}\fi
\ifx\barticle  \undefined \def\barticle#1{}\fi
\ifx\binstitute  \undefined \def\binstitute#1{#1}\fi
\ifx\bpublisher  \undefined \def\bpublisher#1{#1}\fi
\ifx\doiurl    \undefined \def\doiurl#1{\href{#1}{\textsf{DOI}}}\fi
\makeatletter
\def\safeHref#1#2#3{\in@{http}{#2}\ifin@\href{#2}{#3}\else\href{#1#2}{#3}\fi}
\makeatother
\ifx\adsurl    \undefined
  \def\adsurl#1{\safeHref{https://ui.adsabs.harvard.edu/abs/}{#1}{\textsf{ADS}}}\fi
\ifx\arxivurl  \undefined
  \def\arxivurl#1{\safeHref{http://arxiv.org/abs/}{#1}{\textsf{arXiv}}}\fi
\ifx\botherref \undefined \def\botherref#1{}\fi
\ifx\url       \undefined \def\url#1{\textsf{#1}}\fi
\ifx\bchapter  \undefined \def\bchapter#1{}\fi
\ifx\bbook     \undefined \def\bbook#1{}\fi
\ifx\bcomment  \undefined \def\bcomment#1{#1}\fi
\ifx\oauthor   \undefined \def\oauthor#1{#1}\fi
\ifx\citeauthoryear \undefined\def \citeauthoryear#1{#1}\fi
\def\endbibitem {}
\ifx\bconflocation  \undefined \def\bconflocation#1{#1} \fi

\bibitem[\protect\citeauthoryear{{Ahluwalia} and
  {Kamide}}{2004}]{Ahluwalia_2004}
\begin{bchapter}
\bauthor{\bsnm{{Ahluwalia}}, \binits{H.S.}},
\bauthor{\bsnm{{Kamide}}, \binits{Y.}}:
\byear{2004},
\bctitle{{Gnevyshev gap, Forbush decrease, ICME/SSC, and Solar wind}}.
In: \beditor{\bsnm{{Paill{\'e}}}, \binits{J.-P.}} (ed.)
\bbtitle{35th COSPAR Scientific Assembly},
\bfpage{470}.
\adsurl{2004cosp...35..470A}.
\end{bchapter}
\endbibitem

\bibitem[\protect\citeauthoryear{{Badalyan} and
  {Obridko}}{2017}]{Badalyan_2017}
\begin{barticle}
\bauthor{\bsnm{{Badalyan}}, \binits{O.G.}},
\bauthor{\bsnm{{Obridko}}, \binits{V.N.}}:
\byear{2017},
\batitle{{North-south asymmetry of solar activity as a superposition of two
  realizations – the sign and absolute value}}.
\bjtitle{Astron. Astrophys.}
\bvolume{603}.
\doiurl{https://doi.org/10.1051/0004-6361/201527790}.
\end{barticle}
\endbibitem

\bibitem[\protect\citeauthoryear{{Bazilevskaya}, {Makhmutov}, and
  {Sladkova}}{2006}]{Bazilevskaya_2006}
\begin{barticle}
\bauthor{\bsnm{{Bazilevskaya}}, \binits{G.A.}},
\bauthor{\bsnm{{Makhmutov}}, \binits{V.S.}},
\bauthor{\bsnm{{Sladkova}}, \binits{A.I.}}:
\byear{2006},
\batitle{{Gnevyshev gap effects in solar energetic particle activity}}.
\bjtitle{Adv.\ Space\ Res.}
\bvolume{38},
\bfpage{484}.
\doiurl{https://doi.org/10.1016/j.asr.2004.11.011}.
\end{barticle}
\endbibitem

\bibitem[\protect\citeauthoryear{{Carbonell}
  \textit{et~al.}}{2007}]{Carbonell_2007}
\begin{barticle}
\bauthor{\bsnm{{Carbonell}}, \binits{M.}},
\bauthor{\bsnm{{Terradas}}, \binits{J.}},
\bauthor{\bsnm{{Oliver}}, \binits{R.}},
\bauthor{\bsnm{{Ballester}}, \binits{J.L.}}:
\byear{2007},
\batitle{{The statistical significance of the North-South asymmetry of solar
  activity revisited}}.
\bjtitle{Astron. Astrophys.}
\bvolume{476}(\bissue{2}),
\bfpage{951}.
\doiurl{https://doi.org/10.1051/0004-6361:20078004}.
\adsurl{2007A&A...476..951C}.
\end{barticle}
\endbibitem

\bibitem[\protect\citeauthoryear{{Chang}}{2011}]{Chang_2012}
\begin{barticle}
\bauthor{\bsnm{{Chang}}, \binits{H.-Y.}}:
\byear{2011},
\batitle{{Bimodal distribution of area-weighted latitude of sunspots and solar
  North-South asymmetry}}.
\bjtitle{New Astron.}
\bvolume{17}(\bissue{3}),
\bfpage{247}.
\doiurl{https://doi.org/10.1016/j.newast.2011.07.016}.
\adsurl{2012NewA...17..247C}.
\end{barticle}
\endbibitem

\bibitem[\protect\citeauthoryear{{Derrick}, {Deirdre}, and
  {White}}{2016}]{Derrick_2016}
\begin{barticle}
\bauthor{\bsnm{{Derrick}}, \binits{B.}},
\bauthor{\bsnm{{Deirdre}}, \binits{T.}},
\bauthor{\bsnm{{White}}, \binits{P.}}:
\byear{2016},
\batitle{{Why Welch’s test is Type I error robust.}}
\bjtitle{The Quantitative Methods for Psychology}
\bvolume{12}(\bissue{1}),
\bfpage{30}.
\doiurl{https://doi.org/10.1038/367723a0}.
\burl{http://eprints.uwe.ac.uk/27232}.
\end{barticle}
\endbibitem

\bibitem[\protect\citeauthoryear{{Du}}{2015}]{Du_2015}
\begin{barticle}
\bauthor{\bsnm{{Du}}, \binits{Z.L.}}:
\byear{2015},
\batitle{{Bimodal Structure of the Solar Cycle}}.
\bjtitle{Astrophys.\ J.}
\bvolume{804},
\bfpage{15}.
\doiurl{https://doi.org/10.1088/0004-637X/804/1/3}.
\adsurl{2015ApJ...804....3D}.
\end{barticle}
\endbibitem

\bibitem[\protect\citeauthoryear{{Feminella} and
  {Storini}}{1997}]{Feminella_1997}
\begin{barticle}
\bauthor{\bsnm{{Feminella}}, \binits{F.}},
\bauthor{\bsnm{{Storini}}, \binits{M.}}:
\byear{1997},
\batitle{{Large-scale dynamical phenomena during solar activity cycles.}}
\bjtitle{Astron. Astrophys.}
\bvolume{322},
\bfpage{311}.
\adsurl{1997A&A...322..311F}.
\end{barticle}
\endbibitem

\bibitem[\protect\citeauthoryear{{Forbes} \textit{et~al.}}{2011}]{Forbes_2011}
\begin{bbook}
\bauthor{\bsnm{{Forbes}}, \binits{C.}},
\bauthor{\bsnm{{Evans}}, \binits{N.}},
\bauthor{\bsnm{{Hastings}}, \binits{N.}},
\bauthor{\bsnm{{Peacock}}, \binits{B.}}:
\byear{2011},
\bbtitle{{Statistical Distributions}}
\bseriesno{4},
\bpublisher{John Wiley Sons, Inc., Hoboken, New Jersey}, \blocation{???},
\bfpage{47}.
\end{bbook}
\endbibitem

\bibitem[\protect\citeauthoryear{{Gnevyshev}}{1967}]{Gnevyshev_1967}
\begin{barticle}
\bauthor{\bsnm{{Gnevyshev}}, \binits{M.N.}}:
\byear{1967},
\batitle{{On the 11-Years Cycle of Solar Activity}}.
\bjtitle{Sol.\ Phys.}
\bvolume{1},
\bfpage{107}.
\doiurl{https://doi.org/10.1007/BF00150306}.
\adsurl{1967SoPh....1..107G}.
\end{barticle}
\endbibitem

\bibitem[\protect\citeauthoryear{{Gnevyshev}}{1977}]{Gnevyshev_1977}
\begin{barticle}
\bauthor{\bsnm{{Gnevyshev}}, \binits{M.N.}}:
\byear{1977},
\batitle{{Essential features of the 11-year solar cycle}}.
\bjtitle{Sol.\ Phys.}
\bvolume{51},
\bfpage{175}.
\doiurl{https://doi.org/10.1007/BF00240455}.
\adsurl{1977SoPh...51..175G}.
\end{barticle}
\endbibitem

\bibitem[\protect\citeauthoryear{{Ivanov}, {Miletskii}, and
  {Nagovitsyn}}{2011}]{Ivanov_2011}
\begin{barticle}
\bauthor{\bsnm{{Ivanov}}, \binits{V.G.}},
\bauthor{\bsnm{{Miletskii}}, \binits{E.V.}},
\bauthor{\bsnm{{Nagovitsyn}}, \binits{Y.A.}}:
\byear{2011},
\batitle{{Form of the latitude distribution of sunspot activity}}.
\bjtitle{Astron.\ Rep.}
\bvolume{55}(\bissue{10}),
\bfpage{911}.
\doiurl{https://doi.org/10.1134/S1063772911100040}.
\adsurl{2011ARep...55..911I}.
\end{barticle}
\endbibitem

\bibitem[\protect\citeauthoryear{{Javaraiah}}{2012}]{Javaraiah_2012}
\begin{barticle}
\bauthor{\bsnm{{Javaraiah}}, \binits{J.}}:
\byear{2012},
\batitle{{The G-O Rule and Waldmeier Effect in the Variations of the Numbers of
  Large and Small Sunspot Groups}}.
\bjtitle{Sol.\ Phys.}
\bvolume{281},
\bfpage{827}.
\doiurl{https://doi.org/10.1007/s11207-012-0106-6}.
\end{barticle}
\endbibitem

\bibitem[\protect\citeauthoryear{{Javaraiah}}{2016}]{Javaraiah_2016}
\begin{barticle}
\bauthor{\bsnm{{Javaraiah}}, \binits{J.}}:
\byear{2016},
\batitle{{North-south asymmetry in small and large sunspot group activity and
  violation of even-odd solar cycle rule}}.
\bjtitle{Astrophys. Space Sci.}
\bvolume{361},
\bfpage{208}.
\doiurl{https://doi.org/10.1007/s10509-016-2797-x}.
\adsurl{2016Ap\%26SS.361..208J}.
\end{barticle}
\endbibitem

\bibitem[\protect\citeauthoryear{{Krishnamoorthy}}{2006}]{Krishnamoorthy_2006}
\begin{bbook}
\bauthor{\bsnm{{Krishnamoorthy}}, \binits{K.}}:
\byear{2006},
\bbtitle{{Handbook of Statistical Distributions with Applications}},
\bpublisher{Chapman \& Hall/CRC, Taylor \& Francis Group, Boca Raton, FL
  33487-2742}, \blocation{???},
\bfpage{128}.
\bisbn{1-58488-635-8}.
\end{bbook}
\endbibitem

\bibitem[\protect\citeauthoryear{{Leussu} \textit{et~al.}}{2016a}]{Leussu_2016}
\begin{barticle}
\bauthor{\bsnm{{Leussu}}, \binits{R.}},
\bauthor{\bsnm{{Usoskin}}, \binits{I.G.}},
\bauthor{\bsnm{{Arlt}}, \binits{R.}},
\bauthor{\bsnm{{Mursula}}, \binits{K.}}:
\byear{2016}a,
\batitle{{Properties of sunspot cycles and hemispheric wings since the 19th
  century}}.
\bjtitle{Astron. Astrophys.}
\bvolume{592},
\bfpage{A160}.
\doiurl{https://doi.org/10.1051/0004-6361/201628335}.
\adsurl{2016A\%26A...592A.160L}.
\end{barticle}
\endbibitem

\bibitem[\protect\citeauthoryear{{Leussu} \textit{et~al.}}{2016b}]{Leussu_2017}
\begin{botherref}
\oauthor{\bsnm{{Leussu}}, \binits{R.}},
\oauthor{\bsnm{{Usoskin}}, \binits{I.G.}},
\oauthor{\bsnm{{Senthamizh Pavai}}, \binits{V.}},
\oauthor{\bsnm{{Diercke}}, \binits{A.}},
\oauthor{\bsnm{{Arlt}}, \binits{R.}},
\oauthor{\bsnm{{Mursula}}, \binits{K.}}:
2016b,
{VizieR Online Data Catalog: Butterfly diagram wings (Leussu+, 2017)}.
\textit{VizieR Online Data Catalog}
\textbf{359}.
\url{https://vizier.u-strasbg.fr/viz-bin/VizieR?-source=J/A+A/599/A131&-to=3}.
\end{botherref}
\endbibitem

\bibitem[\protect\citeauthoryear{{Li}, {Gao}, and {Zhan}}{2009}]{Li_2009}
\begin{barticle}
\bauthor{\bsnm{{Li}}, \binits{K.J.}},
\bauthor{\bsnm{{Gao}}, \binits{P.X.}},
\bauthor{\bsnm{{Zhan}}, \binits{L.S.}}:
\byear{2009},
\batitle{{The Long-term Behavior of the North South Asymmetry of Sunspot
  Activity}}.
\bjtitle{Sol.\ Phys.}
\bvolume{254}(\bissue{1}),
\bfpage{145}.
\doiurl{https://doi.org/10.1007/s11207-008-9284-7}.
\adsurl{2009SoPh..254..145L}.
\end{barticle}
\endbibitem

\bibitem[\protect\citeauthoryear{{Li} \textit{et~al.}}{2003}]{Li_2003}
\begin{barticle}
\bauthor{\bsnm{{Li}}, \binits{K.J.}},
\bauthor{\bsnm{{Wang}}, \binits{J.X.}},
\bauthor{\bsnm{{Zhan}}, \binits{L.S.}},
\bauthor{\bsnm{{Yun}}, \binits{H.S.}},
\bauthor{\bsnm{{Liang}}, \binits{H.F.}},
\bauthor{\bsnm{{Zhao}}, \binits{H.J.}},
\bauthor{\bsnm{{Gu}}, \binits{X.M.}}:
\byear{2003},
\batitle{{On the Latitudinal Distribution of Sunspot Groups over a Solar
  Cycle}}.
\bjtitle{Sol.\ Phys.}
\bvolume{215}(\bissue{1}),
\bfpage{99}.
\doiurl{https://doi.org/10.1023/A:1024814505979}.
\adsurl{2003SoPh..215...99L}.
\end{barticle}
\endbibitem

\bibitem[\protect\citeauthoryear{{Mandal}, {Karak}, and
  {Banerjee}}{2017}]{Mandal_2017}
\begin{barticle}
\bauthor{\bsnm{{Mandal}}, \binits{S.}},
\bauthor{\bsnm{{Karak}}, \binits{B.B.}},
\bauthor{\bsnm{{Banerjee}}, \binits{D.}}:
\byear{2017},
\batitle{Latitude distribution of sunspots: Analysis using sunspot data and a
  dynamo model}.
\bjtitle{Astrophys.\ J.}
\bvolume{851}(\bissue{1}),
\bfpage{70}.
\doiurl{https://doi.org/10.3847/1538-4357/aa97dc}.
\end{barticle}
\endbibitem

\bibitem[\protect\citeauthoryear{{Munoz-Jaramillo}
  \textit{et~al.}}{2015}]{Munoz-Jaramillo_2015}
\begin{barticle}
\bauthor{\bsnm{{Munoz-Jaramillo}}, \binits{A.}},
\bauthor{\bsnm{{Senkpeil}}, \binits{R.R.}},
\bauthor{\bsnm{{Windmueller}}, \binits{J.C.}},
\bauthor{\bsnm{{Amouzou}}, \binits{E.C.}},
\bauthor{\bsnm{{Longcope}}, \binits{D.W.}},
\bauthor{\bsnm{{Tlatov}}, \binits{A.G.}},
\bauthor{\bsnm{{Nagovitsyn}}, \binits{Y.A.}},
\bauthor{\bsnm{Alexei A.~{Pevtsov}}, \binits{A.A.}},
\bauthor{\bsnm{{Chapman}}, \binits{G.A.}},
\bauthor{\bsnm{{Cookson}}, \binits{A.M.}},
\bauthor{\bsnm{{Yeates}}, \binits{A.R.}},
\bauthor{\bsnm{{Watson}}, \binits{F.T.}},
\bauthor{\bsnm{{Balmaceda}}, \binits{L.A.}},
\bauthor{\bsnm{{DeLuca}}, \binits{E.E.}},
\bauthor{\bsnm{{Martens}}, \binits{P.}}:
\byear{2015},
\batitle{{Small-scale and global dynamos and the area and flux diostributions
  of active regions, sunspot groups, and sunspots: A multi-database study}}.
\bjtitle{Astrophys.\ J.}
\bvolume{800}(\bissue{1}),
\bfpage{48}.
\doiurl{https://doi.org/10.1088/0004-637x/800/1/48}.
\end{barticle}
\endbibitem

\bibitem[\protect\citeauthoryear{{Mursula}, {Usoskin}, and
  {Kovaltsov}}{2001}]{Mursula_2001}
\begin{barticle}
\bauthor{\bsnm{{Mursula}}, \binits{K.}},
\bauthor{\bsnm{{Usoskin}}, \binits{I.G.}},
\bauthor{\bsnm{{Kovaltsov}}, \binits{G.A.}}:
\byear{2001},
\batitle{{Persistent 22-year cycle in sunspot activity: Evidence for a relic
  solar magnetic field}}.
\bjtitle{Sol.\ Phys.}
\bvolume{198}(\bissue{1}),
\bfpage{51}.
\doiurl{https://doi.org/10.1023/A:1005218414790}.
\adsurl{2001SoPh..198...51M}.
\end{barticle}
\endbibitem

\bibitem[\protect\citeauthoryear{{Norton} and {Gallagher}}{2010}]{Norton_2010}
\begin{barticle}
\bauthor{\bsnm{{Norton}}, \binits{A.A.}},
\bauthor{\bsnm{{Gallagher}}, \binits{J.C.}}:
\byear{2010},
\batitle{{Solar-Cycle Characteristics Examined in Separate Hemispheres: Phase,
  Gnevyshev Gap, and Length of Minimum}}.
\bjtitle{Sol.\ Phys.}
\bvolume{261},
\bfpage{193}.
\doiurl{https://doi.org/10.1007/s11207-009-9479-6}.
\adsurl{http://ukads.nottingham.ac.uk/abs/2010SoPh..261..193N}.
\end{barticle}
\endbibitem

\bibitem[\protect\citeauthoryear{{Rice}}{1945}]{Rice_1945}
\begin{barticle}
\bauthor{\bsnm{{Rice}}, \binits{S.O.}}:
\byear{1945},
\batitle{{Mathematical Analysis of Random Noise}}.
\bjtitle{Bell Syst. Techn. J.}
\bvolume{24},
\bfpage{46}.
\end{barticle}
\endbibitem

\bibitem[\protect\citeauthoryear{{Snedecor} and
  {Cochran}}{1989}]{Snedecor_1989}
\begin{bbook}
\bauthor{\bsnm{{Snedecor}}, \binits{G.W.}},
\bauthor{\bsnm{{Cochran}}, \binits{W.G.}}:
\byear{1989},
\bbtitle{{Statistical Methods}},
\bedition{8}th edn.
\bpublisher{Iowa State University Press, Wiley-Blackwell}, \blocation{???},
\bfpage{64}.
\end{bbook}
\endbibitem

\bibitem[\protect\citeauthoryear{{Storini}
  \textit{et~al.}}{2003}]{Storini_2003}
\begin{barticle}
\bauthor{\bsnm{{Storini}}, \binits{M.}},
\bauthor{\bsnm{{Bazilevskaya}}, \binits{G.A.}},
\bauthor{\bsnm{{Fluckiger}}, \binits{E.O.}},
\bauthor{\bsnm{{Krainev}}, \binits{M.B.}},
\bauthor{\bsnm{{Makhmutov}}, \binits{V.S.}},
\bauthor{\bsnm{{Sladkova}}, \binits{A.I.}}:
\byear{2003},
\batitle{{The GNEVYSHEV gap: A review for space weather}}.
\bjtitle{Adv.\ Space\ Res.}
\bvolume{31},
\bfpage{895}.
\doiurl{https://doi.org/10.1016/S0273-1177(02)00789-5}.
\end{barticle}
\endbibitem

\bibitem[\protect\citeauthoryear{{Takalo} and {Mursula}}{2018}]{Takalo_2018}
\begin{barticle}
\bauthor{\bsnm{{Takalo}}, \binits{J.}},
\bauthor{\bsnm{{Mursula}}, \binits{K.}}:
\byear{2018},
\batitle{{Principal component analysis of sunspot cycle shape}}.
\bjtitle{Astron. Astrophys.}
\bvolume{620}.
\doiurl{https://doi.org/10.1051/0004-6361/201833924}.
\end{barticle}
\endbibitem

\bibitem[\protect\citeauthoryear{{Taluktar} and {Lawing}}{1991}]{Taluktar_1991}
\begin{barticle}
\bauthor{\bsnm{{Taluktar}}, \binits{K.K.}},
\bauthor{\bsnm{{Lawing}}, \binits{W.D.}}:
\byear{1991},
\batitle{{Estimation of the parameters of the Rice distribution}}.
\bjtitle{J. Acoust. Soc. Am.}
\bvolume{89}(\bissue{3}),
\bfpage{1191}.
\doiurl{https://doi.org/10.1121/1.400532}.
\end{barticle}
\endbibitem

\bibitem[\protect\citeauthoryear{{Temmer} \textit{et~al.}}{2006}]{Temmer_2006}
\begin{barticle}
\bauthor{\bsnm{{Temmer}}, \binits{M.}},
\bauthor{\bsnm{{Ryb{\'a}k}}, \binits{J.}},
\bauthor{\bsnm{{Bend{\'\i}k}}, \binits{P.}},
\bauthor{\bsnm{{Veronig}}, \binits{A.}},
\bauthor{\bsnm{{Vogler}}, \binits{F.}},
\bauthor{\bsnm{{Otruba}}, \binits{W.}},
\bauthor{\bsnm{{P{\"o}tzi}}, \binits{W.}},
\bauthor{\bsnm{{Hanslmeier}}, \binits{A.}}:
\byear{2006},
\batitle{{Hemispheric sunspot numbers \{R$_{n}$\} and \{R$_{s}$\} from
  1945-2004: catalogue and N-S asymmetry analysis for solar cycles 18-23}}.
\bjtitle{Astron. Astrophys.}
\bvolume{447}(\bissue{2}),
\bfpage{735}.
\doiurl{https://doi.org/10.1051/0004-6361:20054060}.
\adsurl{2006A&A...447..735T}.
\end{barticle}
\endbibitem

\bibitem[\protect\citeauthoryear{{Waldmeier}}{1935}]{Waldmeier_1935}
\begin{barticle}
\bauthor{\bsnm{{Waldmeier}}, \binits{M.}}:
\byear{1935},
\batitle{{Neue Eigenschaften der Sonnenfleckenkurve}}.
\bjtitle{Astron.\ Mitt.\ Zurich}
\bvolume{14},
\bfpage{105}.
\end{barticle}
\endbibitem

\bibitem[\protect\citeauthoryear{{Waldmeier}}{1939}]{Waldmeier_1939}
\begin{barticle}
\bauthor{\bsnm{{Waldmeier}}, \binits{M.}}:
\byear{1939},
\batitle{{Die Zonenwanderung der Sonnenflecken}}.
\bjtitle{Astron.\ Mitt.\ Zurich}
\bvolume{14},
\bfpage{470}.
\adsurl{1939MiZur..14..470W}.
\end{barticle}
\endbibitem

\end{thebibliography}
%
% Without BibTeX 
% \begin{thebibliography}{}
% \bibitem[\protect\citeauthoryear{Author}{Year}]{key}
%   <bibliographical entry>
%
% \bibitem[\protect\citeauthoryear{}{}]{}
%   
%  
% \end{thebibliography}

\end{article} 
\end{document}